\documentclass[10pt,letterpaper]{article}
\usepackage{lscape}
\usepackage[top=0.85in,left=2.5in,footskip=0.75in]{geometry}

\raggedright
\usepackage{amsmath,amssymb}

\usepackage{changepage}

\usepackage{textcomp,marvosym}

\usepackage{cite}

\usepackage{nameref,hyperref}

\usepackage[right]{lineno}

\usepackage[nopatch=equm]{microtype}
\DisableLigatures[f]{encoding = *, family = * }

\usepackage[table]{xcolor}

\usepackage{array}

\newcolumntype{+}{!{\vrule width 2pt}}

\newlength\savedwidth

\usepackage[aboveskip=1pt,labelfont=bf,labelsep=period,justification=raggedright,singlelinecheck=off]{caption}

\makeatletter
\renewcommand{\@biblabel}[1]{\quad#1.}
\makeatother

\usepackage{lastpage,fancyhdr,graphicx}
\usepackage{epstopdf}
\fancyheadoffset[L]{2.25in}
\fancyfootoffset[L]{2.25in}
\newcommand\widebar{\overline} 

\newcommand{\vect}{\boldsymbol } 
\newcommand{\est}{\widehat}
\newcommand{\mean}{\widebar}

\newcommand{\dspaceoff}{\renewcommand{\baselinestretch}{1}
            \large\normalsize}

\newcommand{\cit}[5]{$_{ {#1}\ \!}{#2}_{ #3_{\ } }  ( {#4 }){ \ #5 }$}

\begin{document}
\vspace*{0.2in}

\begin{flushleft}
{\Large
\textbf\newline{
Rank-based  methods for estimating landmark win probability in longitudinal randomized controlled trials with missing data} 
}
\newline
\\
Guangyong Zou \textsuperscript{1,2  \ddag }\textsuperscript {*},
Shi-Fang Qiu \textsuperscript{3 \ddag},
Joshua Zou \textsuperscript{4 },
Emma Davies Smith \textsuperscript{5},
Yun-Hee Choi \textsuperscript{1 },
Yuhan Bi \textsuperscript{1 } 
\\
\bigskip
\textbf{1} Department of Epidemiology and Biostatistics, Schulich School of Medicine \& Dentistry, Western University, London, Ontario, Canada
\\
\textbf{2} Alimentiv Inc, London,  Ontario, Canada
\\
\textbf{3} Department of Statistics and Data Science, Chongqing University of Technology, Chongqing  China
\\
\textbf{4} Department of Statistics and Actuarial Science, University of Waterloo, Waterloo, Canada 
\\
\textbf{5} Department of Biostatistics, Harvard T H Chan School of Public Health, Boston, Masschusetts, USA

\bigskip

%
%
\ddag These authors contributed equally to this work.





* gzou2@uwo.ca

\end{flushleft}
\section*{Abstract}
 The primary analysis for longitudinal randomized controlled trials (RCTs) often compares treatment groups at the last timepoint,  referred to as the landmark time. 
Assuming data are normally distributed and missing at random, the mixed model for repeated measures (MMRM) is widely used  to conduct inference   in terms of a mean difference.
When outcomes violate normality assumption and/or the mean difference lacks a clear interpretation, we may quantify treatment effects using the  probability   that a treated participant would have a better outcome than (or win over) a control participant. For RCTs with missing data, one may apply the generalized pairwise comparison (GPC) procedure, which carries forward the results of a pairwise  comparison from a  previous timepoint.
We propose first  using ranks to  converts each observation at a timepoint into   a win fraction, reflecting the proportion of times that the observation is better than every observation in the comparison group. Then, we conduct inference for  the win probability  based on the win fractions using the MMRM to obtain the point and variance estimates. 
Simulation results suggest that  our method  performed much better than  the GPC procedure. We illustrate our proposed procedure in SAS and R using data from two published trials.




\section{Introduction}\label{sec-intro}

Treatment effects are typically assessed using randomized controlled trials (RCTs) in which each participant is followed  over time and outcome variables are measured at several pre-specified timepoints \cite{malli2003,molen2004,malli2008}. The primary analysis usually compares the outcome data between different treatment groups at the end of follow up \cite{bell2016}, which may be  referred to as the landmark \cite{mallinckrodt2019}. When the outcome variable possesses   meaningful units and approximately follows a normal distribution, the treatment effect may be quantified by a point estimate of the mean difference between groups at the landmark, accompanied by the associated confidence interval and $P$-value.

Missing data is an ever-present problem. Some participants may drop out or be lost during the follow-up  before the completion of the study, resulting in missing data for the landmark analysis.   As classified by  Rubin \cite{rubin1976}, data are missing completely at random (MCAR) if, conditional on the covariates,  the probability of outcome missing  is unrelated to either the observed or unobserved outcomes; missing at random (MAR) when missingness does not depend on the unobserved data, conditional on the observed data; and missing not at random (MNAR) when missingness depends on  unobserved data even after conditioning on the observed data. 

The assumption of MAR is commonly made for the primary analysis of RCTs \cite{ little2012prevention}. A likelihood-based mixed-effects model for repeated measures (MMRM) with a group-specific  ‘unstructured’ covariance matrix has been recommended for analyzing longitudinal continuous outcomes collected at fixed time points \cite{malli2003,malli2008}. The MMRM has  several advantages. First, it   is  robust to  model misspecification because of the unstructured time-response profiles. Second, it allows for the use of partial data from participants who dropped out during the trial and do not have data for the landmark timepoint. Third, the MMRM with a group-specific ‘unstructured’  covariance matrix for   within-subject correlations is robust to misspecification of the covariance structure and variance heterogeneity. Finally, it can be more efficient than multiple imputation (MI) in many situations, despite the latter being commonly regarded as the gold standard approach for handling missing data \cite{sullivan2018}. In terms of the estimands framework outlined in the ICH E9(R1) Addendum, the  MMRM analysis commonly uses the hypothetical strategy to handle dropouts by  estimating the treatment effect that would have been observed   had there been no dropouts \cite{mallinckrodt2019}.

Besides missing data, longitudinal RCTs often employ outcome measures that do not follow normality and/or lack meaningful units, such as disease severity scores or health related quality of life scores  \cite {mallinckrodt2019,ratitch2020}.
Despite its popularity   in RCTs, less attention has been given to the analysis of such data.

 Without making distributional assumptions for outcome data, we may quantify the treatment effect withthe results of pairwise comparisons, an idea that dates back to Deuchler in 1914  \cite{kruskal57}. Specifically, Deuchler assigned a score of +1, -1, or 0 to each pair of observations from the comparison groups, when the first observation of the pair is greater than, less than, or equal to the second observation. A statistic  was then constructed using the sum of  scores divided by the number of pairs with nonzero scores. This idea has been re-discovered and modified many times in the literature, most notably by Mann and Whitney  \cite{mann1947}.  
 
Buyse  \cite{buyse2010} 
 used the term  generalized pairwise comparison (GPC) to describe the above scoring system, and net treatment benefit to describe the resulting statistic.  
The  net treatment benefit, also known as the Mann-Whitney difference \cite{lachin1992} or Somers' $D$ \cite{newson2002},  is   the difference between the probability that a randomly selected observation from  the treatment group is better than (or wins over) a randomly selected observation from  the control  group
and the probability that a randomly selected observation from the control group wins over a randomly selected observation from the treatment group.  
Since these two  win probabilities (WinPs) sum to one, we need only  focus on  inference of the WinP by the treatment group.

One novel aspect of the Buyse GPC  procedure   in the context of longitudinal RCTs is that   timepoints can be prioritized according to clinical importance, which is also the  principle underlies the win ratio analysis \cite{pocock2024}. 
Thus, it seems very attractive to apply the GPC procedure for landmark analysis  in longitudinal RCTs with missing data.
 Specifically, the timepoints from baseline can be ordered in reverse such that the landmark timepoint becomes  the most important one,  and the second from the last becomes the second important one, etc.  If a   treatment-control pair cannot be scored due to missing data at the highest priority (primary) timepoint, then a comparison can be made using   the next timepoint in the hierarchy. Note that this approach has also appeared in the literature under the  terms  `mean rank imputation’ \cite{huang2013, fan2016}, and ‘last kernel carried forward’ \cite{sun2017}, and has been  built into the R-package {\tt sanon} \cite{sanon2015}.

We previously developed methods for the design and analysis of RCTs with a single followup time \cite{zou2022,zou2023a,zou2023b}.
The   objective of this manuscript is to extend our previous  regression approach to estimating the WinP  in longitudinal RCTs with missing data.   Our method first converts each observation at each timepoint to a win fraction that represents  the fraction of times that the observation is better  than all the observations in the comparison group. 
Using the relationship between ranks  and the Deuchler scoring system, i.e, the sign of a difference between two numbers \cite{kruskal57,mann1947,buyse2010},  the win fractions can be easily obtained based on ranks. 
We then use  regression models with dependent variable being  the win fractions to obtain the point and variance estimates, followed by invoking the central limit theorem for inference.  One advantage of this  approach is that it can be easily implemented using standard statistical software.

 The remainder of the manuscript is structured as follows. Section 2 presents two classic datasets as motivating examples. Section 3  first presents the regression  approach to estimating WinP in the absence of missing data, followed by the GPC procedure, complete case analysis (ignoring cases with missing timepoint data), and MMRM for win fractions analysis.
    Section 4 presents  a simulation study to evaluate the three methods. Section 5 illustrates  the methods using data arising from the motivating examples.  We close the manuscript with a discussion.   We present  illustrative SAS and R code in the online Appendix.

\section{Two Motivating Examples}

\subsection {The postnatal depression trial}
 
\begin{figure}
  
 \begin{center}
 
\includegraphics[width=\linewidth]{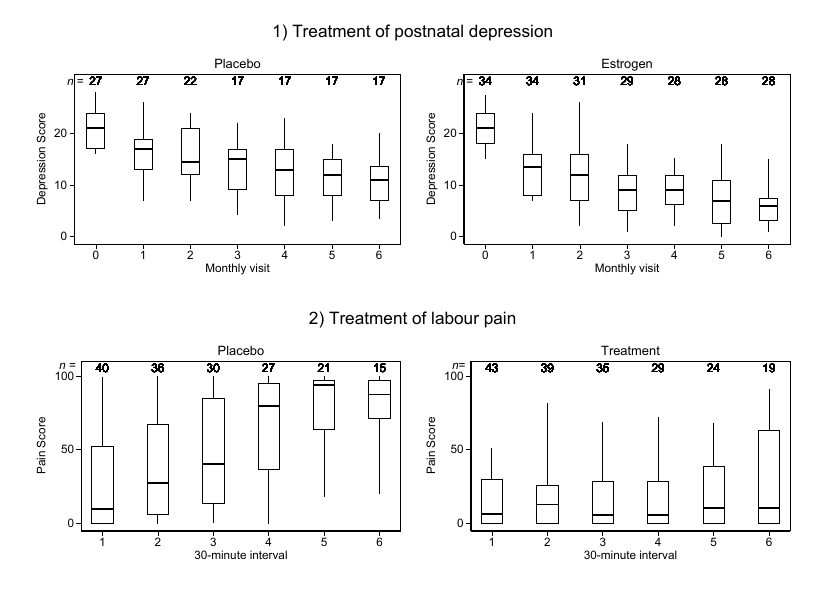}

\end{center}
 \caption{Boxplots by treatment group: 1) the Edinburgh postnatal depression scale (EPDS) scores at    baseline (visit 0) and six post-intervention visits in a randomization trial evaluating  an estrogen patch for treating postnatal women with major depression  \cite{gregoire96} and 2) the six 30-minute intervals pain scores  in a trial evaluating a treatment for maternal pain relief during labor \cite{davis1991}.  $n$ denotes the number of participants at each timepoint.
     }\label{everitt_dep}

\end{figure}

 This trial investigated the effectiveness of   oestrogen when taken transdermally by woman diagnosed with postpartum depression  \cite{gregoire96}.  A total of 61 participants were randomized to either   an oestradiol ($n = 34$), two at a time, to receive a total daily dose of $200 \mu g$ of $17\beta$-oestradiol, or a placebo patch ($n = 27$). Each subject was evaluated on the Edinburgh postnatal depression scale (EPDS) at baseline and throughout the six monthly visits. To complete the EPDS questionnaire,  respondents     selected the number next to the response that best corresponded to how they  felt during the past seven days. For example, possible responses to the statement  `The thought of harming myself has occurred to me' were 3 (Yes, quite often), 2 (Sometimes), 1 (Hardly ever), and 0 (Never).
The total score was obtained by summing the numbers selected for each of the 10 items, 
 with a maximum score of 30 points indicating severe depression.  The data are available in  Rabe-Hesketh and Everitt \cite{everittbk}, pp. 138-139.  The data is also reproduced in the supplementary SAS code for convenience. According to  these authors \cite{everittbk},  the non-integer scores resulted from substituting the average of all available items for  the  missing questionnaire items. 
The visit-specific boxplots by treatment group are presented in Figure \ref{everitt_dep}, showing that by visit 6,  18\% of participants dropped out in the control group and 37\% in the treatment group.
 One could apply the MMRM directly to the EPDS scores, resulting in inference on the mean difference in scores, which may not be easily interpreted.

\subsection{The trial of labor pain}
 Davis \cite{davis1991} presented data from a randomized trial evaluating a treatment for maternal pain relief during labor. In this trial, 43 women in labor were randomly assigned to receive an experimental pain medication and 40 women were assigned to receive a placebo.
Treatment was initiated when the cervical dilation was 8 cm. At 30 minutes intervals, the intensity of pain was self-reported by placing a mark on a 100 mm line (0=no pain, 100=very much pain). As shown in Figure \ref{everitt_dep}, the data are very skewed and have numerous missing values at later time intervals.  By the 6th time interval,   62.5\% of participants  in the placebo and 55.8\% in the treatment group  dropped out of the study.
There has been no easy way to analyze such data to estimate the effect of  the drug in reducing labor pain.

\section {Methods}

\subsection {Notations,  definitions, and estimators}

 Consider a longitudinal randomized  clinical trial with a two-group parallel design. Suppose that the primary objective is to compare the outcome at the end of the treatment period by estimating the treatment effect.   
 Let $\mathbf{Y}_{ij} =[Y_{ij0},Y_{ij1}, \dots, Y_{ijT}]^{T}$ denote a vector of observations for the $j$th  participant in the $i$th group, with $Y_{ijt}$ representing  measurements at timepoint $t$, $t=0, 1, 2, \dots, T$, with $t=0$ denoting baseline, and $i=0$  denoting the control group and $i=1$ the treatment group, where $j=1, 2, \dots, n_i$, and total sample size is $N=n_1+ n_0$. 
    We denote  the sample sizes in group $i$ at timepoint $t$ as $n_{it}$. When there is no missing data, $n_{it}=n_i$ for all $t=0, 1, 2, \dots, T$.
 
 Data for different subjects are assumed to be independent. However, repeated observations from the same subject are assumed to be correlated with a   cumulative distribution function (CDF) for outcome at time $t$ given by
 \[F_{it}(y) = \frac{1}{2}\big[F_{it}^-(y) + F_{it}^+(y)\big]
 \]
  where $F_{it}^-(t)=\Pr(Y_{ijt} <y)$ is the left-continuous version and $F_{it}^+(y)=\Pr(Y_{ijt} \leq y)$ is the right-continuous version of the CDF. Such a definition can easily handle ties \cite{brunner2002}. An estimator for $F_{it}(y)$ is given by
  \[
  \est F_{it}(y) = \frac{1}{n_{it}} \sum_{j=1}^{n_{it}} H(y, Y_{ijt})
  \]
  where $H(a,b)$ is defined via the Deuchler score, or the sign of $a-b$,  as 
  \[
  H(a,b) = \frac{1}{2}\big[\mbox{sign}(a-b)+1\big]
\] 
    
   We can now define the distribution-free treatment effect at time $t$ as the probability that an observation  in the treatment $Y_{1jt}$, $j=1, 2, \dots, n_1$,  is larger   than (or wins over) an observation in the control $Y_{0j't}$, $j'=1, 2, \dots, n_0$. 
 Then  the win probability (WinP) at timepoint $t$ can be defined as 
  \begin{equation}\label{def}
           \theta_{t} = P(Y_{0j't} < Y_{1jt}) + \frac{1}{2}P(Y_{0j't} = Y_{1jt}) =\int_{}^{}F_{0t}dF_{1t},
            \quad t =0, 1,2,\dots,T
  \end{equation}
  and  estimated as
  \[
\est     \theta_{t} = \int_{}^{} \est F_{0t}d \est F_{1t}
  \]
        An analogous definition can be given for situations where smaller values are regarded as `winning'.  Many  terms have appeared in the literature for $\theta_{t}$, including the area under the receiver operating characteristic curve \cite{bamber75,delong1988,hanley1982}, ridits \cite{bross1958,beder1990}
   and the $c$-statistic \cite{harrell1996}. We prefer the term WinP for its intuitive interpretation.

 For normally distributed data with a homogeneous variance between the two treatment groups, i.e. $Y_{ijt} \sim N(\mu_{it}, \sigma_t^2)$, the WinP is given by
$
	\theta_{t} = \Phi\Big[ ( \mu_{1t} - \mu_{0t})/(\sqrt{2} \sigma_{t})\Big]
$,
where $(\mu_{1t} - \mu_{0t})/\sigma_{t}$  is commonly known as the standardized mean difference (SMD) or Cohen's effect size  \cite{cohen92}.
Thus, inference for the SMD can be conducted using a WinP analysis, thereby providing a viable alternative  approach to constructing a confidence interval for the SMD. 
This relationship also suggests that Cohen's popular benchmarks for the SMD may be adapted for  interpreting the WinP. That is, WinPs of 0.56, 0.64, and 0.71 correspond to  SMDs of 0.2 `small', 0.5 `medium', and 0.8 `large' effect size, respectively. 

In addition,  the net treatment benefit defined by  Buyse \cite{buyse2010} is given by
$ \mbox{NB}_t =  P(Y_{0j't} < Y_{1jt}) -  P(Y_{0j't} > Y_{1jt}) = 2\theta_t -1$, which   can be  recognized as the   risk difference for binary outcomes.
The win odds (WO)   \cite {dong2020}, previously known as the generalized odds ratio \cite{agresti80},  is also readily available as 
$\mbox{WO}_t =  \theta_t/(1-\theta_t)$. Thus, the null value of the WinP,  $\theta_t =0.5$, corresponds to NB$_t$ = 0 and WO$_t$ = 1.
 We focus on developing  methods for the WinP in longitudinal RCTs, 
  however the results are readily applicable to the NB,  WO, and the SMD without making distributional assumptions for the outcome data.

\subsection{Inference for a single win probability with complete data}

To conduct inference, we assume $\min[n_{1t}, n_{0t}]\rightarrow \infty$ such that $N/n_{it} < \infty$.  The distribution of $\sqrt{N} (\est \theta_t -\theta_t)$ can be established by first deriving the point estimator and variance for WinP, followed by invoking the central limit theorem. Following Brunner and Munzel   \cite{brunner2000},  
 \begin{equation}\label{decomp}
  \begin{array}{ll}  
\sqrt{N} (\est \theta_t -\theta_t) & = \displaystyle \sqrt{N} \Big\{ \int \est F_{0t} \mbox{d} \est F_{1t} - \int F_{0t} \mbox{d} F_{1t}\Big\} \\
 & =\displaystyle \sqrt{N} \Big\{ \int F_{0t} \mbox{ d}(\est F_{1t} - F_{1t}) + \int (\est F_{0t} - F_{0t}) \mbox{ d} F_{1t} + \int (\est F_{0t} - F_{0t})  \mbox{ d}(\est F_{1t} - F_{1t}) \Big\} \\
   & \approx   \displaystyle \sqrt{N} \Big\{  \frac{1}{n_{1t}} \sum_{j=1}^{n_{1t}} \Big[F_{0t}(Y_{1jt}) -\theta_t\Big] - \frac{1}{n_{0t}}\sum_{j'=1}^{n_{0t}} \Big[F_{1t}(Y_{0j't}) - \big(1- \theta_t \big)\Big]  \Big\}      \\
   & =  \displaystyle \sqrt{N} \Big\{  \frac{1}{n_{1t}} \sum_{j=1}^{n_{1t}} F_{0t}(Y_{1jt}) - \frac{1}{n_{0t}}\sum_{j'=1}^{n_{0t}} F_{1t}(Y_{0j't}) + 1 - 2\theta_t
    \Big\}  
 \end{array}
 \end{equation}
 where the approximation holds due to  $\displaystyle\sqrt{N} \int (\est F_{0t} - F_{0t})  \mbox{ d}(\est F_{1t} - F_{1t})\rightarrow 0$  in probability.
 This suggests that
\begin{equation}\label{equiv}
\sqrt{N} (\est \theta_t -\theta_t)  \stackrel{d}{\approx} \sqrt{N} \Big\{  \frac{1}{n_{1t}} \sum_{j=1}^{n_{1t}} F_{0t}(Y_{1jt}) - \frac{1}{n_{0t}}\sum_{j'=1}^{n_{0t}} F_{1t}(Y_{0j't}) + 1 - 2\theta_t     \Big\}  
    \end{equation}
where $\stackrel{d}{\approx}$ denotes asymptotic equivalence in distribution. Three consequences follow from Eq (\ref{equiv}).
 First, since the expectation of the right side  of Eq (\ref{equiv}) is  0, we have 
   \begin{equation}\label{winpestp}
   \theta_t = \frac{1}{2} \Big[\frac{1}{n_{1t}}   \sum_{j=1}^{n_{1t}}     F_{0t} (Y_{1jt})   -  \frac{1}{n_{0t}}\sum_{j'=1}^{n_{0t}}    F_{1t} (Y_{0j't})\Big] +0.5
   \end{equation}
 Second, the distributional equivalence of the two sides in Eq (\ref{equiv}) suggests
 \begin{equation}\label{v1}
 \mbox{var}(\est\theta_t) \approx  \mbox{var}\Big[\frac{1}{n_{1t}}   \sum_{j=1}^{n_{1t}}     F_{0t} (Y_{1jt})   -  \frac{1}{n_{0t}}\sum_{j'=1}^{n_{0t}}    F_{1t} (Y_{0j't})\Big]
 \end{equation}
 Finally,  an application of the central limit theorem to the right side  yields that   \cite{brunner2000}
 \begin{equation}\label{clt1}
 \frac{ \est\theta_t -\theta_t }{\sqrt{ \mbox{var}(\est\theta_t)}}  \sim N(0,1)
 \end{equation}
Inference for $\theta_t$ based on the observed data  can proceed by plugging-in the  empirical CDFs.
Specifically,   the empirical CDFs can be written as
\begin{equation}\label{winf}
\est F_{0t} (Y_{1jt}) = \frac{1}{n_{0t}} \sum_{j'=1}^{n_{0t}} H(Y_{1jt}, Y_{0j't})  
\quad \mbox{ and }\quad
\est F_{1t} (Y_{0j't}) = \frac{1}{n_{1t}} \sum_{j=1}^{n_{1t}} H(Y_{0j't}, Y_{1jt})
\end{equation}
 which are consistent estimators for $F_{0t}(Y_{1jt})$ and $ F_{1t} (Y_{0j't})$, respectively.
   Note that  $\est F_{0t} (Y_{1jt})$ quantifies the proportion of times that an observation in the treatment group, $Y_{1jt}$, wins over all observations in the control group, while 
  $\est F_{1t} (Y_{0j't})$ quantifies the proportion of times that an observation in the control group, $Y_{0j't}$, wins over all observations in the treatment group. 
  The win fractions for the treatment group and one minus the win  fractions in the control group  have been termed as `placements' \cite{hanley1997,dodd2003}.
  
    For  simplicity,    we let
   \[
   W_{1jt}  =\est F_{0t}(Y_{1jt}) \quad{\mbox{and}} \quad   W_{0j't}  =   \est F_{1t}(Y_{0j't})
   \]  
   with means defined as
   \[
  \mean W_{i.t} = \frac{\sum_{j=1}^{n_{it}} W_{ijt}}{n_{it}}, \quad  i=0, 1  
   \]
   Thus, the estimators based on  Eq (\ref{winpestp})  may be obtained as
   \begin{equation}\label{winpest}
   \est\theta_t = \mean W_{1.t} = \frac{1}{2} (\mean W_{1.t} - \mean W_{0.t}) +0.5
   \end{equation} 
 with the corresponding variance estimator based on Eq (\ref{v1})  given by    
  \begin{equation}\label{vare}
  \est{\mbox{var}}(\est\theta_t) = \est{\mbox{var}}\Big[\frac{1}{n_{1t}}   \sum_{j=1}^{n_{1t}}  W_{1jt} -
  \frac{1}{n_{0t}}\sum_{j'=1}^{n_{0t}}  W_{0j't}  \Big] = \est{\mbox{var}}(\mean W_{1.t}-    \mean W_{0.t})
  \end{equation}
     Note that   the ostensible mismatch between  $\est\theta_t$ and its variance  may also be explained by the fact that   $W_{ijt}$ is constructed by conditioning on the outcome variable of the comparison group. 
   
To simplify the calculation of  win fractions, we make the link between win fractions and ranks, defined via the Deuchler scores. 
  At timepoint $t$, the rank of an observation in the treatment group, $Y_{1jt}$, among all observation in the combined sample of $n_{1t} + n_{0t}$ is given by  \cite{hoeffding1948},
\[
	R_{1jt} = \frac{1}{2} +   \sum_{j'=1}^{n_{1t}} H(Y_{1jt},Y_{1j't}) + \sum_{j'=1}^{n_{0t}}H(Y_{1jt},Y_{0j't})  
\]
and the within-group rank is given by,
\[
	r_{1jt} = \frac{1}{2} + \sum_{j'=1}^{n_{1t}} H(Y_{1jt},Y_{1j't})
\]
Thus,
\[ 
	 W_{1jt} =\frac{R_{1jt} - r_{1jt}}{ n_{0t}}   = \frac{ \sum_{j'=1}^{n_{0t}} H(Y_{1jt},Y_{0j't})}{ n_{0t}}   \quad  \quad j=1, 2, \dots, n_{1t}
\]
quantifying the proportion of times that an observation wins over every observation in the control group.
Similarly, the win fractions for observations in the control group, $Y_{0j't}$ can be obtained as
\[
	W_{0j't} = \frac{R_{0j't} - r_{0j't}}{ n_{1t}}, \quad j'=1, 2, \dots, n_{0t} 
\]

 To  obtain the point estimate and variance of the WinP  in Eqs (\ref{winpest}) and (\ref{vare}),     we  use the readily available regression procedures in common statistical software.       Specifically, we   regress   the win fractions, $W_{ijt}$, on the treatment group indicator (0 for control and 1 for treatment) and other covariates (or their win fractions).    Let  $G_{ij} = i$ be the treatment group indicator, we fit the following model
 \begin{equation}\label{regsinglew}
 W_{ijt} = \beta_0 + \beta_1 G_{ij}  +e_{ijt}
 \end{equation} 
 which yields $\est\beta_1 = \mean W_{1.t} -    \mean W_{0.t}$. Thus by Eq (\ref{winpest})
  \[
  \est \theta_t = \frac{\est  \beta_1}{2} +0.5
  \]
  Furthermore, by Eq (\ref{vare})
   \[
 \est{\mbox{var}}(\est\theta_t)  =  \est{\mbox{var}}( \mean W_{1.t} -    \mean W_{0.t})  =    \est{\mbox{var}}( \est\beta_1) 
  \] 
We emphasize that the regression model is used a tool to obtain estimates of the WinP and  variance simultaneously, although they are derived from two different consequences of the asymptotic equivalence in distribution of  Eq (\ref{equiv}).
 Thus,  it is important not to apply conventional results    
      from regression models, leading to the erroneous conclusion  that   $\est{\mbox{var}}(\est \theta)= \est{\mbox{var}}(\est\beta_1)/4$.

 We do not assume variance homogeneity in this framework as it is well-known to be too restrictive  for rank-based analysis \cite{brunner2002,rubarth2022}.
    Instead,  we apply the  robust variance estimator \cite{mackinnon1985}, yielding
  \begin{equation}\label{sandi}
    \est{\mbox{var}}( \est\beta_1)  =  \frac{1}{n_{1t}(1-n_{1t})} \sum_{j=1}^{n_{1t}}\big(W_{1jt}-\mean W_{1.t})^2 + \frac{1}{n_{0t}(1-n_{0t})} \sum_{j'=1}^{n_{0t}}\big(W_{0j't}-\mean W_{0.t})^2
  \end{equation}
which is identical to variance estimators that have appeared repeatedly in the literature  \cite{sen1967,delong1988,brunner2000}.

To see how to obtain Eq (\ref{sandi}) using the simple linear regression in Eq (\ref{regsinglew}) with the sandwich variance estimator approach \cite{mackinnon1985}, denote the design matrix $\vect G$ as 
\[
G = \vect 1_n || (\vect 1_{n_1} //\vect 0_{n_0})
\]
where $ \vect 1_n$ is a vector of 1's and $\vect 0_{n_0}$ is a vector of 0's, 
then the `robust' or `sandwich' variance estimator for $\est \beta_1$ is the (2,2) element of 
\[
\vect V = \big(\vect G^T\vect G\big)^{-1} \vect G^T \mbox{diag}\left(\frac{e_{ijt}^2}{1-h_{ij} }\right) \vect G \big(\vect G^T\vect G\big)^{-1}
\]
where 
\[
\big(\vect G^T\vect G\big)^{-1} =\begin{bmatrix}
1/n_0 & -1/n_0\\
-1/n_0 & 1/n_1 + 1/n_0\\
\end{bmatrix},
\quad h_{ij} =\begin{bmatrix} 1 & G_{ij}\end{bmatrix} 
\big(\vect G^T\vect G\big)^{-1}
\begin{bmatrix} 1 \\ G_{ij}\end{bmatrix}
\] 
Thus, the (2,2) element of $\vect V$
  is given by
\[
\est{\mbox{var}}(\est\beta_1) =  \frac{\sum_{j=1}^{n_1} e_{1jt}^2 }{n_1(n_1-1)} +  \frac{\sum_{j=1}^{n_0} e_{0jt}^2 }{n_0(n_0-1)}
\]
which can be obtained using standard statistical procedures, e.g.,  SAS {\tt PROC MIXED} with option {\tt GROUP=} treatment indicator in the {\tt REPEATED} statement.  
   Note also that  the normality assumption for the residuals is not critical as we do not conduct inference for $\beta_1$  with a $t$-distribution.
 Rather, we only use regression procedures that are readily available in software packages  as a device to obtain the estimators in Eqs (\ref{winpest}) and (\ref{vare}).

    As pointed out by Sen  \cite{sen1967}, the   variance estimator in Eq (\ref{sandi}) is positively biased, with a  bias less than $\theta_t (1-\theta_t)/N$.  Tcheuko et al \cite{tcheuko2016} provided a practical approach to compute the unbiased variance. Brunner and Konietschke \cite{brunner2024} also proposed an unbiased variance estimator. Both estimators are  most useful for   RCTs with small sample sizes, e.g., $N< 30$, which is not our focus, and will not be pursued further here.  We do not pursue these estimators further here  since it is not obvious to us how to implement them in the regression framework so that more general cases can be handled.
  
In what  follows,  we  apply regression models to   $W_{ijt}$ as dependent variables, and then  use Eqs (\ref{winpest}) and (\ref{vare}) to  identify  estimates of  $\theta_t$ and the  corresponding variance. Inference on WinP proceeds by applying the Slutzky theorem to Eq (\ref{clt1}), yielding asymptotically
 \begin{equation}\label{clt}  \frac{\est\theta_t -\theta_t}{\sqrt{\est{\mbox{var}}(\est\theta_t)}}  \sim N(0,1) 
  \end{equation}

 To improve small sample performance,  
 we   conduct inference on the logit-scale of the WinP,  i.e., ln(win odds). Specifically, for testing $H_0: \theta_t=0.50$, with $\est{\mbox{se}}(\est\theta_t)$ being the square root of $\est{\mbox{var}}(\est\theta_t)$,  the test statistic is given by
\begin{equation}\label{test}
T = \frac{\ln \big[\est\theta_t/(1-\est\theta_t)\big]}{
\est {\mbox{se}}(\est\theta_t)/\big[\est\theta_t(1-\est\theta_t) \big]
 }
\end{equation}
which is distributed asymptotically as the standard normal. The corresponding $(1-\alpha)100\%$ confidence interval for $\theta_t$ is given by
\begin{equation}\label{winpci}
\frac{\exp(l)}{1+\exp(l)}   \quad \mbox{  to }\quad   \frac{\exp(u)}{1+\exp(u)}
\end{equation}
where the confidence limits for ln(win odds) are given by
\[
l, u= \ln \frac{\est\theta_t }{1- \est\theta_t }\pm z_{\alpha/2} \frac{\est{\mbox{se}}(\est\theta_t)}{\est\theta_t(1- \est\theta_t) } 
\] 
and $z_{\alpha/2}$ is the upper $\alpha/2$ quantile of the standard normal distribution.

\subsection{The generalized pairwise comparison procedure  for WinP with missing data}\label{GPC}
 
 In the context of longitudinal RCTs, the  GPC procedure \cite{buyse2010}  starts by ordering  timepoints according to clinical importance. Usually the later timepoints are considered to be  more clinically important than the earlier ones. The WinP at the last timepoint $T$, $\theta_T$, may be estimated using the two-sample $U$-statistic as
\begin{equation}\label{thetat}
	\est \theta_{T} = \frac{1}{n_1 n_0}\sum_{j=1}^{n_1}\sum_{j'=1}^{n_0} H(Y_{1jT},Y_{0j'T})
\end{equation}
when both $Y_{1jT}$ and $Y_{0j'T}$ are observed. When either $Y_{1jT}$ or $Y_{0j'T}$ or both are missing, the GPC procedure  replaces  $H(Y_{1jT}, Y_{0j'T})$  with $H(Y_{1j(T-1)},Y_{0j'(T-1)})$ if both $Y_{1j(T-1)}$ and $Y_{0j'(T-1)}$ are observed. Otherwise, it is replaced with $H(Y_{1j(T-2)},Y_{0j'(T-2)})$. This process continues until a score is assigned for each pair of $Y_{1jt}$ and $Y_{0j't}$ for all $j = 1,\dots, n_1$ and $j' = 1, \dots, n_{0}$. 
Note that   the sample size is still $n_i$ since the GPCs carries   forward the comparisons prior to the end of treatment period.  
In light of results by Rauch et al \cite{rauch2014}, the estimated win probability using the GPC procedure may  be seen  as a weighted average of the timepoint-specific win probabilities, with non-standardized weights ranging from 0.5 to 1. 
Non-standardized GPC weights cast doubt on the validity and interpretation of estimates, and their complex form has also been discussed by others\cite{zhou2022,fuyama2023}.   
 
A previous simulation study 
found that  the GPC procedure can inflate Type I error even when data are missing under MCAR in the case of a single follow-up timepoint \cite{thomas2023}, while  another study \cite{fan2016} provided supporting evidence for the use of the GPC procedure for hypothesis testing.    Neither of these papers have discussed estimation and confidence interval construction  for  appropriate treatment effects.

 The estimator for $\theta_T$ can be re-written as 
\[
	\begin{array}{ll}
	\est \theta_{T} & =\displaystyle  \frac{1}{n_1}\sum_{j=1}^{n_1}\bigg[\frac{1}{n_0}\sum_{j'=1}^{n_0} H(Y_{1jT},Y_{0j'T}) \bigg] \\
	& = \displaystyle \frac{1}{n_1}\sum_{j=1}^{n_1} W_{1jT} = \mean W_{1.T}
	\end{array}
\] 
where 
\[
W_{1jT} = \frac{1}{n_0}\sum_{j'=1}^{n_0} H(Y_{1jT},Y_{0j'T}), \quad j=1, 2, \dots, n_1
\]
 which represents the fraction of times that an observation in group 1 at timepoint $T$, $Y_{1jT}$, wins over every observation in group 0 at timepoint $T$,  $Y_{0j'T}$, for $j=1, 2, \dots, n_1$ and $j'=1, 2, \dots, n_0$.  
 The win fraction for an observation in the  group 0, $Y_{0j'T}$, can be obtained analogously as 
\[
W_{0j'T} = \frac{1}{n_1}\sum_{j=1}^{n_1} H(Y_{0j'T},Y_{1jT}), \quad j'=1, 2, \dots, n_0 
\]

With the constructed $W_{ijt}$, we   use a regression model similar to  Eq (\ref{regsinglew}) to obtain the point estimate for  $\theta_T$  and its variance, with the guidance of Eqs (\ref{winpest}) and (\ref{vare}). Specifically, by fitting $W_{ijT}$ to the following model 
 \begin{equation}\label{regsingleww}
 W_{ijT} = \beta_0 + \beta_1 G_{ij} +e_{ijT}
 \end{equation} 
 we obtain 
  \[
\est  \theta_T =  \frac{ \est \beta_1}{2} + 0.5
  \]
  and
\begin{equation}\label{regsinglev}
 \begin{array}{ll} 
  \est{\mbox{var}}(\est\theta_T)  & = \est{\mbox{var}}(\est\beta_1) =\est{\mbox{var}}(\mean W_{1.T} - \mean W_{0.T}) \\ 
   &  = \displaystyle \frac{1} {n_1}s_{1T}^2  + \frac{1}{n_0}s_{0T}^2 
 \end{array}
 \end{equation}
 with
 \[
 s_{iT}^2 = \frac{1}{n_i-1} \sum_{j=1}^{n_i} \big(W_{ijT} - \mean W_{i.T} \big)^2, \quad i=1, 0
 \]  

To follow the analysis of adjusting for baseline outcome measurements to improve power in RCTs, we can extend the model to an analysis of covariance (ANCOVA)-type model as  \cite{zou2023a,zou2023b}
\begin{equation}\label{gpc}
W_{ijT} = \beta_{0} + \beta_1 G_{ij} +\gamma W_{ij0} +  e_{ijT} 
\end{equation}
where $W_{ij0}$ denotes win fraction for the baseline outcome measurements and  $e_{ijT}\sim N(0, \sigma_i^2)$.  
With estimates of  $\est\theta_T$ and its variance, we can conduct inference   on the logit-scale of WinP,  i.e., ln(win odds) as in Eqs (\ref{test}) and (\ref{winpci}).

 Note that we have used the  win fractions, $W_{ijT}$, as a simple and flexible alternative  to the  current implementation of the GPC procedure  \cite{buyse2010,jaspers2024}. 
Specifically, the   GPC analysis involves $n_1\times n_0$ data points with three possible values of 1, 0.5, and -1, while our procedure analyzes $n_1+ n_0$ data points with possible values ranging from 0 to 1.

\subsection{The mixed-effects model regression approach for win probability}
	
	Instead of carrying the last comparison forward as done by the GPC approach, we can apply  the MMRM for timepoint-specific win fractions, resulting in a similar nonparametric approach to multivariate outcomes of different scales \cite[]{zouzou2024}.  
	Recall that the timepoint-specific win probability is given by 
\[
	\begin{aligned}
\hat\theta_{t} & = \frac{1}{n_{1t}n_{0t}}\sum_{j=1}^{n_{1t}}\sum_{j'=1}^{n_{0t}}H(Y_{1jt},Y_{0j't}) \\
& = \frac{1}{n_{1t}}  \sum_{j=1}^{n_{1t}}\bigg[\frac{1}{n_{0t}}\sum_{j'=1}^{n_{0t}}H(Y_{1jt},Y_{0j't})\bigg]
 = 1 - \frac{1}{n_{0t}} \sum_{j'=1}^{n_{0t}}\bigg[\frac{1}{n_{1t}}\sum_{j=1}^{n_{1t}}H(Y_{0j't},Y_{1jt}) \bigg]\\
& = \frac{1}{n_{1t}}\sum_{j=1}^{n_{1t}}W_{1jt} 
= 1 - \frac{1}{n_{0t}}\sum_{j'=1}^{n_{0t}}W_{0j't} 
	\end{aligned}
\]
where
\[
 W_{1jt} = \frac{1}{n_{0t}}\sum_{j'=1}^{n_{0t}} H(Y_{1jt},Y_{0j't}) \mbox{ and }   W_{0j't} = \frac{1}{n_{1t}}\sum_{j=1}^{n_{1t}} H(Y_{0j't},Y_{1jt}) 
 \]

As discussed above, the time-specific win fractions,  $W_{ijt}$,  can be conveniently calculated using (mid)ranks. In the context of disease severity, we use descending ranks.  

The simplest approach is to ignore cases with missing data and analyze the win fractions, $W_{ijT}$, at the  last timepoint, using an ANCOVA model given by 
\begin{equation}\label{cca}
	W_{ijT} = \beta_{0} + \beta_{1M} G_{ij}+ \gamma W_{ij0} + e_{ijT} 
\end{equation}
where $W_{ij0}$ represents the win fractions based on baseline measurements,  $\beta_{1M}$ is the   difference in mean fractions between two treatment groups at time $T$, and  $e_{ijT}\sim N(0, \sigma_i^2)$. Note that this model has the same form as in a  complete case analysis (CCA), but the win fractions here are constructed using timepoint-specific ranks.

Again, based on Eq (\ref{winpest}),   the corresponding WinP at time $T$ is estimated by
\[
\est\theta_T =\frac{\est\beta_{1M} }{2} +0.5
\] 
and by Eq  (\ref{vare}) the variance is estimated by
\[
\est{\mbox{var}}(\est\theta_T)  \approx \est{\mbox{var}}(\mean W_{1.T} - \mean W_{0.T}) = \est{\mbox{var}}(\est\beta_{1M})
\]

Inference for the WinP can proceed on the logit scale using Eqs (\ref{test}) and (\ref{winpci}). 
We refer to this approach as the  CCA.
A key limitation of CCA is that it does not use all available data, and thus may lose information under MCAR, or may provide biased results under  MAR or MNAR \cite{malli2008,siddiqui2011mmrm}.

 To overcome such limitation, we may analyze the win fractions, $W_{ijt}$, with a mixed model for repeated measures (MMRM) \cite{mall2001}. Denoting $\vect W_{ij} =[W_{ij1}, W_{ij2}, \dots, W_{ijT}]^T$, the model is given by,
\begin{equation}\label{mmrm}
	\vect{W}_{ij} = \vect{X}_{ij} \vect \beta  + \vect{e}_{ij} 
\end{equation}
where
\[
	\vect{X}_{ij} =   \big[1,  \vect{G}_{ij},   W_{ij0} \big] \bigotimes \vect{I}_T   \mbox{, and }\vect{\beta} =[\beta_{01}, \dots, \beta_{0T},  \beta_{11}, \dots, \beta_{1T}, \gamma_1, \gamma_2, \dots, \gamma_T]^T
\]
with  $\bigotimes$ denoting the Kronecker product,  $\vect I_T$ being a $T\times T$ identity matrix, and the superscript  `$T$' denoting the transpose, and $\vect e_{ij} \sim N(\vect 0, \vect \Sigma_i)$, with an unstructured (co)variance matrix $\vect \Sigma_i$ for group $i$.

Missing outcome data can be easily handled by removing the corresponding rows in $\vect X_{ij}$ and elements in $\vect \beta$.
 When coding in statistical software, $\vect{X}_{ij}$ may be defined by regarding   the variable for timepoints   as a categorical factor   crossed with the treatment indicator and baseline win fractions, $W_{ij0}$. Separate unstructured covariance structures for each group $\vect \Sigma_i$ can be easily implemented using the {\tt GROUP} option of the {\tt REPEATED} statement in {\tt PROC MIXED} with SAS.
Point estimates of all timepoint-specific WinPs, including $\theta_T$,  and associated standard errors can be obtained using least-squares mean estimates (`lsmestimates') of differences  from  the MMRM model.

Since this model uses an unstructured time profile for fixed effects and covariance structure for each treatment group, it avoids model misspecification  and provides valid results for data under MCAR or MAR in the analysis of longitudinal RCTs with continuous data \cite{malli2008,siddiqui2011mmrm}.
Again, the performance of this method may be improved by a  logit-transformation for $\est\theta_T$ by adapting Eqs (\ref{test}) and (\ref{winpci}).
The finite sample performance of MMRM for analyzing win fractions  is largely unknown and will be investigated by simulation study below.

\section {Simulation Study}

Since the three methods (the GPC in Eq (\ref{gpc}), CCA in Eq (\ref{cca} ) and MMRM in Eq (\ref{mmrm})) presented  in the previous section were developed using large sample theory, their performance in finite samples must be evaluated. To this end, we evaluate the performance of these methods in terms of the relative  bias \% for estimating the WinP ($\theta_T$) at the pre-specified final  timepoint, the empirical coverage and tail errors of the associated two-sided 95\% confidence intervals, as well as the rejection rates in testing $H_{0} : \theta_T = 0.5$. 

The three methods provided identical results when there are no missing data and no baseline measurements. Thus, we focused on cases of missing data with adjustments for baseline measurements. 
We simulated complete data from a multivariate normal distribution with parameter values patterned based on the mean depression scores at Baseline, and Visits 2, 4, and 6 and variance-covariance matrices  in the postpartum depression  study \cite{gregoire96}. 

We considered four trajectories.  In Trajectory 1, both group means changed over time, but were equal at each timepoint. In Trajectory 2, both group means changed over time, but group mean for the treatment arm changed more before converging to the same mean of the control arm at the last timepoint. In Trajectory  3, the treatment-arm mean deviated  more over time,   representing a `small' effect size of $\theta_T \approx 0.56$.  In Trajectory 4, the profiles were crossed with the  win probability at the last timepoint representing close to a `medium' effect size of  $\theta_T \approx 0.62$.
The specific timepoint mean vectors for the control and treatment arms $\vect m_0, \vect m_1$   for each trajectory are given by
\begin{enumerate}
\item[1)] $\vect m_0= \vect m_1 = (20, 16, 12, 11)$
\item[2)] $\vect m_0=(20, 16, 12, 11) $ and $\vect m_1 =(20, 15, 9, 11)$
\item[3)] $\vect m_0=(20, 16, 12, 11) $ and $\vect m_1 =(20, 15, 9, 10)$
\item[4)] $\vect m_0=(20, 15, 9, 11) $ and $\vect m_1 =(20, 16, 12, 9)$
\end{enumerate}
 and the variance-covariance matrices for the two groups  are  given by 
 \[ 
\vect \Sigma_0 =
\begin{bmatrix}
  15.6 & 12.9 &  4.8 &  4.4\\
12.9  & 37.5 &  22.8 & 11.6\\
 4.8  & 22.8 & 34.2 &  17.9\\
 4.4  & 11.6 &  17.9 &  21.9\\
\end{bmatrix} 
  \mbox{ and }
\vect \Sigma_1= \begin{bmatrix}
12.8  & 7.4 &   3.6 &   7.1 \\
 7.4  & 43.2 &  22.7 &  23.8 \\ 
 3.6  & 22.7 & 21.8 & 18.8 \\
 7.1  & 23.8 & 18.8 &  22.4  \\    
\end{bmatrix}
\]
for the control and treatment  groups, respectively.

Since lower scores indicate better outcomes, for the  rank-based  procedure, we used reverse ranking in the simulation, i.e.,  the largest score receives a rank of 1, the next largest score receives a rank of 2, and so on. For the GPC procedure, we used $H(a,b)=[1-\mbox{sign}(a-b)]/2$ for calculating win fractions.

We first considered a sample size of $n_{1} = n_{0} = 50$, similar to our motivating examples. 
We repeated the above process for each scenario 1000 times. 
The relative  bias \% is defined as  the mean of  $(\est{\rm WinP} - {\rm WinP})/{\rm WinP}\times 100$ over 1000 replications.  
Performance of the 95\% confidence interval is quantified by overall coverage, defined as the percentage of times the confidence interval covers the true parameter value,  and the left- and right-tail  errors,  defined as    the percentage of times the upper limit is smaller   and the lower limit is greater  than the true parameter value, respectively. We considered adequate empirical coverage to be within 93.6\% to 96.4\%. 
We quantified the  empirical power  as the percentage of times that the null hypothesis $H_0:$ WinP = 0.5 was rejected at the 2-sided 5\% significance level.  
We also calculated the mean width of the 1000 confidence intervals for each scenario.

For MCAR, we deleted data such that 10\% of participants dropout at Times 2, 3, and 4, respectively, resulting in a  30\% total dropout rate at Time 4  for both treatment groups. 

For MAR and MNAR, we adapted the methods used by Mallinckrodt et al  \cite{mallinckrodt2004} and  Barnes et al \cite{barnes2008}  to simulate two-arm trials with outcomes measured at baseline and three post-intervention timepoints. Specifically, we deleted  the complete data   for each of the four trajectories according to the following four combinations:
	\begin{itemize}
		\item[(1)] Equal trigger values ($>$16) and equal dropout probability (0.4) for both groups;
		\item[(2)] Different trigger values ($>$16 for group 0 and $>$15 for group 1) and equal dropout probability (0.4);
		\item[(3)] Equal trigger values ($>$16) and different dropout probabilities for both groups (0.5 for group 0 and 0.3 for group 1); and
		\item[(4)] Different trigger values ($>$16 for group 0 and $>$15 for group 1) and different dropout probabilities for both groups (0.5 for group 0 and 0.3 for group group 1).
	\end{itemize}
To create data with a MAR dropout mechanism, scores that triggered dropout were retained, so that the observed data could explain the dropout. For example, if a participant had a score of 17 at Time 2 and  the realized value of Bernoulli (0.4) is 1, then 17 is the trigger for dropout. For MAR, the value of 17 is retained, but the values at Times 3 and 4 were deleted. To create data with a MNAR dropout mechanism, scores that triggered dropouts were also deleted, so that the observed data could not entirely explain the dropout. In the above example, the value at Time 2 is also deleted.

The resulting dropout rates by treatment group under MAR and MNAR are presented given in Table \ref{simtable}. 
Each dataset was then analyzed using  the GPC, CCA, and MMRM procedures, with the latter two approaches analyzing the win fractions converted using ranks.  For comparison, we also analyzed each dataset prior to creating missing data.  

Table \ref{mcar} shows the results of the four trajectories with a 30\% dropout rate under MCAR.
The results  clearly show that the GPC method only performed well under Trajectory 1, i.e, the null hypothesis is true under all the time points, which is very restrictive in practice.
In particular, the  bias was less than 5\% for both CCA and MMRM methods, while the GPC exhibited bias of 57.5\%, -41.2\%, and -170.7\% for Trajectories 2 to 4, respectively. Consequently, the CI coverage rates for the CCA and MMRM are close to the nominal level of 95\%, while the coverage for the GPC was far from  95\%. 
 Interestingly, the GPC method provided narrower confidence intervals than the datasets without missing data.
This property has been documented for last observation carried forward (LOCF) in the literature \cite{molen2004,malli2008,beunckens2005}. 
Under the other three trajectories, the GPC method may overestimate or underestimate the true parameter values of the win probability depending on the magnitudes of separation prior to the primary timepoint. The confidence intervals failed to maintain the nominal coverage level. These results reinforce the conclusion arrived by Deltuvaite-Thomas and Burzykowski \cite{thomas2023} that the GPC method should not be used in practice with missing data, despite recommendations by   Byuse \cite{buyse2010} and Fan et al  \cite{fan2016}.
In fact, our simulation results demonstrate similar characteristics of the    LOCF  technique, which cannot provide valid results even when the data are MCAR  \cite{beunckens2005,siddiqui2009}.


Table \ref{traj12} presents the results for Trajectory 1 (Cases 1 to 4) and Trajectory 2 (Cases 5 to 8) under the four combinations of data deletion to emulate data  MAR and MNAR. 
Under the MAR settings, the overall coverage percentages of confidence intervals from the GPC for Trajectory 1, i.e., the null hypothesis is true at all timepoints, are close to the nominal level of 95\%. However, the confidence intervals are lopsided, with unequal tail errors. For Trajectory 2, the GPC method  resulted in biased point estimates, poor coverage percentages and inflated Type I error rates. The CCA method performed better than the GPC method, but was still unsatisfactory. The MMRM for win fractions performed very  well for all eight scenarios in terms of bias, coverage percentage, and Type I error rate. Under the MNAR settings, neither the GPC method nor the CCA method can be recommended for practice. Interestingly, the MMRM procedure performed reasonably well, suggesting that it is fairly robust to violation of MAR assumption. This is  consistent with  previous results when applying the  MMRM to conduct inference on mean scores \cite{mallinckrodt2004,barnes2008,siddiqui2009}.

Table \ref{traj34} shows the results of the true win probability greater than 0.5 in Trajectory 3 (Cases 9 to 12) and Trajectory 4 (cases 13 to 16) under the four combinations of data deletion to emulate data  MAR and MNAR. 
For Trajectory 3, the GPC method had comparable confidence interval coverage rates to the MMRM that are close to the nominal level. For Trajectory 4, the GPC produced  bias\%  ranging from -58.2\% to -92.1\%, with confidence interval coverage rates severely below the nominal level. Although the CCA performed better than the GPC under this trajectory, it did not   perform  as well as the MMRM, which provided reasonable results across all criteria.
Overall,   the MMRM provided satisfactory results, and was   fairly robust to the violations of the MAR assumption.

Simulation results (not shown) for sample sizes  $n_0=50$ and $n_1=100$ and $n_0=100$ and $n_1=100$ are consistent with the above conclusions. In addition, unreported results also suggest that the performance of the methods for log-normal data are  identical to the above results, which is expected because all methods are rank-based and invariant to any monotone transformation.

\section{Analyzing the motivating examples}

We now apply the GPC, CCA and MMRM methods to data from the two motivating examples, with 
the primary goal  of estimating the win probability at the end of the treatment period, in addition to time-specific WinP from the MMRM procedure.  Since lower scores indicate better outcomes (wins) in both examples, for the rank-based procedures, we ranked the data in descending order as done in the simulation. For the GPC procedure, we used $H(a.b)=[1-\mbox{sign}(a-b)]/2$ to calculate win fractions.

For the postnatal depression trial \cite{gregoire96},  the visit-specific boxplots show no strong evidence that the normality assumption is  violated.  
One could apply the MMRM on the raw scores, resulting in $P= 0.0026$ for testing the mean score difference between the treatment and control groups, and point estimate and 95\% CI given by 4.685 (1.757, 7.613).  Except for  the significant $P$-value, the point estimate and CI are not straightforward to interpret.
 
Table \ref{example} presents the results of the WinP analysis using the three approaches. The estimate at Visit 6 (95\% CI) based on the GPC method  
 is given by  0.737 (0.611, 0.834) with $P=0.0005$, compared with the CCA method point estimate of 0.779 (0.604, 0.890). The corresponding results based on the MMRM for timepoint-specific win fractions
are 0.777 (0.608, 0.887) and $P=0.0025$. In light of the simulation results, it is reasonable to conclude that the GPC method provided biased results. We can also explain the results by regarding the GPC estimate as a weighted average of the timepoint-specific WinPs, since the visit-specific estimates before Visit 6 are smaller in this case. 
In other words, carrying forward these smaller  pairwise-comparisons by the GPC resulted in underestimates for the landmark WinP.  
The similar results from the CCA and MMRM methods suggest that the MAR assumption is tenable for this example, and the results from the MMRM are reliable. The results may be reported as ``the probability that a patient with depression treated with a daily dose of $200 \mu g$ of $17\beta$-oestradiol had a better EPDS score than a control patient is  77.7\% (95\% CI 60.8 to 88.7\%, $P=0.0025$)\rq\rq{}. 

For the labor pain trial \cite{davis1991},   Figure \ref{everitt_dep} clearly demonstrates that the data    are far from normal.  In addition, there are substantial dropouts at later time intervals. 
We thus refrain from applying the MMRM directly to the pain scores.    
Table \ref{example} presents the results of the WinP analysis.
The estimate (95\% CI) of win probability  at the sixth time interval based on the GPC method is 0.756 (0.649, 0.838) with $P=0.00005$. The CCA method yielded the estimate of 0.895 (0.738, 0.962) and $P=0.00014$. The results are comparable with those of the MMRM for win fractions, with point estimate (95\% CI) at the sixth time-interval given by 0.875 (0.722, 0.950) and $P=0.00012$. 
 In conclusion, we report the probability that a woman treated with pain medicine had less pain than a  woman in the control arm  is 87.5\% (95\% CI 72.2 to 95.0\%, $P=0.00012$). 
Again, the underestimation by the GPC method can be explained by examining the   interval-specific estimates from the MMRM for win fractions.

The superficially narrower confidence intervals by the GPC method  for both examples are the result of treating the comparisons carried forward for missing data as real data. This feature of masking the uncertainty of missing data has been well-documented in the literature for LOCF \cite {molen2004,beunckens2005}.  
 The LOCF treats the imputed values and the observed value on equal footing, while the GPC method  treats the imputed comparisons and actual comparisons on equal footing and thus artificially increases the amount of available information. The simulation results also demonstrate this  characteristic of the GPC method.

\section{Discussion}

Longitudinal RCTs are commonly analyzed by estimating treatment effect at a pre-specified  landmark timepoint.  Missing data due to dropout is usually handled with the hypothetical strategy so that the estimand targets  treatment effect under the assumption that `patients take their medication as directed' \cite {mallinckrodt2019}.  

When the outcome measurements  are assumed to be normally distributed  and inference is on the difference between group means, there exists a large literature \cite{malli2003,molen2004,malli2008,siddiqui2011mmrm,siddiqui2009}.
When outcome data does not follow normal distribution or  lacks clearly meaningful units, the MMRM approach may  not be directly applicable. In such situations,   the GPC procedure appears to be an attractive   option due to its ability to prioritize timepoints \cite{buyse2010}.  
 Despite the  large literature on the GPC procedure in RCTs, especially in situations where different types of outcomes  are involved in the construction of   a composite end point \cite{gpcbook,pocock2024},  its validity for handling missing data in landmark analysis has not been fully explored, with few exceptions \cite{fan2016,thomas2023}, which  provided   conflicting results. 
 
We found that the GPC method can yield biased results even under restrictive assumptions, such as data being missing completely at random and equal dropout rates between the two comparison groups. 
The GPC method appears to be valid only when there is no treatment effects at all timepoints. 
This result is important, since    the GPC method was suggested by Buyse \cite{buyse2010} as an alternative to the LOCF approach, which   is well-known to lead to  misleading inference   \cite {molen2004,lachin2016}.   
The simulation results showed that the GPC method has the similar problem associated with  LOCF. This is not surprising because  the only difference between the two approaches is that the GPC carries last comparison forward, whereas LOCF carries the last observation forward.  
In fact, the GPC approach in the present context results in a point estimate that may be regarded as a    weighted average of timepoint-specific win probabilities, with complicated non-standardized weights that depend on the correlations among the repeated measures and dropout rates as well as ties \cite{rauch2014}.  Similar problems  have been identified  for the infamous  LOCF approach \cite{molen2004,lachin2016}.

  
We proposed a rank-based simple alternative that   involves three steps. First, convert the timepoint-specific raw scores to win fractions using ranks. Second, analyze the win fractions using the MMRM with least square contrasts to obtain estimates of the timepoint-specific win probabilities and their standard errors. Finally, apply the logit-transformation for inference in terms of the win probability for the primary timepoint and other timepoints as required.   The simulation results suggest our approach   performed very well in terms of bias and confidence interval coverage, as long as the data are missing under MCAR or MAR. The MMRM procedure for win fractions has also demonstrated some robust properties to data under MNAR. Similar to the conventional MMRM analysis, our approach is useful for primary analysis when the hypothetical strategy is used to deal with dropouts.

 Our approach in principle is similar to the nonparametric approach by   Rubarth et al \cite{rubarth2022}, who developed a procedure for  comparing multiple groups based on ranks by also invoking the `Asymptotic Equivalence Theorem' given by Eq (\ref{equiv}). However, their approach is more  suitable for controlled or homogeneous study environments, requiring  no adjustment for covariates.
As a regression approach, our approach can  readily handle  covariates \cite{zou2023a,zou2023b}, even in the case of correlated or clustered outcomes \cite{zou21,zouzou2024,davies2025}.  
   Moreover,  the WinP estimates from our approach are collapsible due to the use of a linear model for estimation. Thus, our procedure makes moot the debate of which estimand to use in RCTs with discrete data.

We did not consider multiple imputation (MI).  Previous simulation   results   \cite{thomas2023} showed that MI may result in severely conservative Type I error rates across a variety of scenarios to the extent that the empirical rejection rates are around 2.5\% for the nominal 5\% level, as shown in their online Supplementary Tables S13 to S16. 
Second, MI usually relies on parametric imputation models, including the predictive mean matching method. It is unclear to us on how to proceed with ordered categorical data or severely skewed outcomes as those in the labour pain example. Addressing these difficulties to exploit  the advantage of MI in using the information contained in the auxiliary variables  is left for future research. 
  
In summary, we have identified the limitations of using the   GPC  method for handling missing data in longitudinal trials when the primary comparison is at the  last planned timepoint (landmark).
The MMRM procedure for win fractions, which can be obtained easily using ranks,  presents a valid and much simpler alternative for estimating win probability. The estimated win probability  can be directly transformed to net treatment  benefit \cite{buyse2010,lachin1992} and win odds \cite{dong2020}. Due to unstructured time and covariance structure, the MMRM procedure for win fractions can be pre-specified in the protocol and the statistical analysis plan  \cite{malli2008,mallinckrodt2019}. In the protocol development, the defining attributes of the estimands remain the same as those for longitudinal RCTs using the  hypothetical strategy to deal with missing data as detailed by Mallinckrodt et al \cite{mallinckrodt2019}, except the `population summary' is win probability rather than `mean difference'. 
As a rank-based procedure, our approach   requires no user-written specialized software packages. 
The SAS and R code for analyzing the postnatal depression trial is presented in the (online) Appendix.

\section {Acknowledgments} 
The research of Drs G Zou and Choi was supported partially by Individual Discovery Grants from the Natural and Engineering Research Council of Canada, Grant/Award Number: RGPIN-2019-04741, RGPIN-2019-06549. Dr. Qiu is supported by the Science and Technology Research Program of Chongqing Municipal Education Commission (Grant No. KJZD-K202201101).

\section{Appendix (online)}

\subsection{SAS code}
In this appendix, we present SAS and R code for the analysis of the postnatal depression trial (EPDS) to estimate win probability using the MMRM for win fractions. Note that the non-integer scores resulted from substituting the average of all available items for the missing questionnaire items. Three steps  are involved. First, Convert wide format data to timepoint-specific win fractions using PROC RANK, with the option DESCENDING to accommodate smaller scores win. Second, Analyze win fractions in long format using PROC MIXED and obtain timepoint-specific win probability through treatment-by-time interaction contrasts using LSMESTIMATE statements. Finally, Manipulate results from LSMESTIMATE with logit-transformation to obtain the win probability estimates and confidence interval for each timepoint.

 \dspaceoff
\begin{verbatim} 
**EPDS Data;
data wideEPDS;
input id trt y0 y1 y2 y3 y4 y5 y6@@;
cards;
1 0 18 17 18 15 17 14 15 2 0 27 26 23 18 17 12 10
3 0 16 17 14 . . . . 4 0 17 14 23 17 13 12 12
5 0 15 12 10 8 4 5 5 6 0 20 19 11.54 9 8 6.82 5.05 
7 0 16 13 13 9 7 8 7 8 0 28 26 27 . . . .
9 0 28 26 24 19 13.94 11 9 10 0 25 9 12 15 12 13 20
11 0 24 14 . . . . . 12 0 16 19 13 14 23 15 11
13 0 26 13 22 . . . . 14 0 21 7 13 . . . .
15 0 21 18 . . . . . 16 0 22 18 . . . . .
17 0 26 19 13 22 12 18 13 18 0 19 19 7 8 2 5 6
19 0 22 20 15 20 17 15 13.73 20 0 16 7 8 12 10 10 12
21 0 21 19 18 16 13 16 15 22 0 20 16 21 17 21 16 18
23 0 17 15 . . . . . 24 0 22 20 21 17 14 14 10
25 0 19 16 19 . . . . 26 0 21 7 4 4.19 4.73 3.03 3.45
27 0 18 19 . . . . . 28 1 21 13 12 9 9 13 6
29 1 27 8 17 15 7 5 7 30 1 15 8 12.27 10 10 6 5.96
31 1 24 14 14 13 12 18 15 32 1 15 15 16 11 14 12 8
33 1 17 9 5 3 6 0 2 34 1 20 7 7 7 12 9 6
35 1 18 8 1 1 2 0 1 36 1 28 11 7 3 2 2 2
37 1 21 7 8 6 6.5 4.64 4.97 38 1 18 8 6 4 11 7 6
39 1 27.46 22 27 24 22 24 23 40 1 19 14 12 15 12 9 6
41 1 20 13 10 7 9 11 11 42 1 16 17 26 . . . .
43 1 21 19 9 9 12 5 7 44 1 23 11 7 5 8 2 3
45 1 23 16 13 . . . . 46 1 24 16 15 11 11 11 11
47 1 25 20 18 16 9 10 6 48 1 22 15 17.57 12 9 8 6.5
49 1 20 7 2 1 0 0 2 50 1 20 12.13 8 6 3 2 3
51 1 25 15 24 18 15.19 13 12.32 52 1 18 17 6 2 2 0 1
53 1 26 1 18 10 13 12 10 54 1 20 27 13 9 8 4 5
55 1 17 20 10 8.89 8.49 7.02 6.79 56 1 22 12 . . . . . 
57 1 22 15.38 2 4 6 3 3 58 1 23 11 9 10 8 7 4
59 1 17 15 . . . . . 60 1 22 7 12 15 . . . 
61 1 26 24 . . . . . 
; 
*First, create win fractions; 
proc sort data=wideEPDS; by trt; run;
ods listing close;
ods output summary= NN(keep=trt y0_N y1_N y2_N y3_N y4_N y5_N y6_N );
proc means data= wideEPDS; BY trt;
var y0 -y6; run; 
data NN; *switch sample size at each visit;
set NN; trt = 1-trt;
run; 
proc sort data=NN; by trt; run; 
proc rank descending data= wideEPDS out=overrank (keep=yo0 - yo6 trt id); 
var y0-y6; 
ranks yo0 - yo6; run; 
proc rank descending data=wideEPDS out=grprank (keep=yg0 -yg6 trt id);
by trt;
var y0 -y6; 
ranks yg0 -yg6; run;
data widewinF; *win fractions;
merge overrank grprank NN; by trt; 
y0 = (yo0 - yg0)/y0_N;
y1 = (yo1 - yg1)/y1_N;
y2 = (yo2 - yg2)/y2_N;
y3 = (yo3 - yg3)/y3_N;
y4 = (yo4 - yg4)/y4_N;
y5 = (yo5 - yg5)/y5_N;
y6 = (yo6 - yg6)/y6_N;
keep trt id y0- y6; run; 
data longWinF;
set widewinF;
time='y1'; winF = y1; output;
time='y2'; winF = y2; output; 
time='y3'; winF = y3; output;
time='y4'; winF = y4; output; 
time='y5'; winF = y5; output; 
time='y6'; winF = y6; output; 
keep id trt time winF y0; run; 
proc sort data = longWinF out= longWF; by outcome ; run;
*Second, analyze long fortmat win fraction;
ods listing close; 
ods output LSMEstimates = MMRMEst;
proc mixed data = longWF; 
class id trt outcome;
model winF =y0*time  time*trt/noint notest ddfm =kr; 
repeated time /subject =id type =un group=trt; 
lsmestimate trt*time [-1, 1 1][1, 2 1]; 
lsmestimate trt*time [-1, 1 2][1, 2 2]; 
lsmestimate trt*time [-1, 1 3][1, 2 3]; 
lsmestimate trt*time [-1, 1 4][1, 2 4]; 
lsmestimate trt*time [-1, 1 5][1, 2 5]; 
lsmestimate trt*time [-1, 1 6][1, 2 6]; run; 
*Finally, obtatin WINP estimated and 95% CIs; 
data WinP;
set MMRMEst;
WinP = Estimate/2+.5; 
** lgt transformation works better; 
lgt =log(WinP /(1-WinP )); 
selgt = StdErr/(WinP*(1-WinP)); 
l = lgt - 1.96*selgt;
u = lgt + 1.96*selgt;
low = logistic(l );
upp = logistic(u );
wid=upp-low; 
test=lgt/selgt;
p_val = 2*(1-probnorm(abs(test))) ; 
keep winP low upp wid p_val; run; 
ods listing;
proc print ; 
var winP low upp wid p_val; run; 
\end{verbatim}

\subsection{R code} 
 
 \begin{verbatim}
 # Load libraries
library(dplyr)
library(tidyr)
library(mmrm)
library(emmeans)

epds_wide <- read.csv("wideEPDS.csv")

# Transform to long format
epds_long <- epds_wide |> 
pivot_longer(!c(id, trt, y0), names_to = "t", values_to = "y") |>
mutate(time = as.numeric(gsub("y", "", t))) |>
select(-t)

# Transform time, id and trt to factor format
epds_long$time <- as.factor(epds_long$time)
epds_long$id <- as.factor(epds_long$id)
epds_long$trt<- as.factor(epds_long$trt)

######--------------transform to rank based win fractions-----------------####
# Calculate sample size based on different trt and time(excluding missing)
epds_long2 <- epds_wide |> 
pivot_longer(!c(id, trt), names_to = "t", values_to = "y") |>
mutate(time = as.numeric(gsub("y", "", t))) |>
select(-t)

# Switch n and count by group and time
count_data <- epds_long2 |>
mutate(trt = ifelse(trt == "0", "1", "0")) |>
group_by(trt, time) |> 
summarize(n = sum(!is.na(y))) |> 
ungroup() 

# Create ranks(descending)
ranks <- epds_wide |>
mutate(R0 = rank(-y0, na.last = "keep",),
R1 = rank(-y1, na.last = "keep"),
R2 = rank(-y2, na.last = "keep"),
R3 = rank(-y3, na.last = "keep"),
R4 = rank(-y4, na.last = "keep"),
R5 = rank(-y5, na.last = "keep"),
R6 = rank(-y6, na.last = "keep")) |>
group_by(trt) |>
mutate(r0 = rank(-y0, na.last = "keep"),
r1 = rank(-y1, na.last = "keep"),
r2 = rank(-y2, na.last = "keep"),
r3 = rank(-y3, na.last = "keep"),
r4 = rank(-y4, na.last = "keep"),
r5 = rank(-y5, na.last = "keep"),
r6 = rank(-y6, na.last = "keep")) |>
mutate()

# Transform ranks to the long format
ranks_long <- ranks |> 
pivot_longer(cols = c(y0:y6, R0:R6, r0:r6), 
names_to = c(".value", "time"), 
names_pattern = "(.)(.)") %>% 
mutate(time = as.numeric(time)) |>
merge(count_data, by = c("trt", "time")) |>
 #merge with count data n by group and time
mutate(dd = R -r,
winf = dd/n,
time = as.factor(time),
trt = as.factor(trt),
id = as.factor(id)) #calculate the group difference and win fractions

# Get the column winf0 for baseline
winf0 <- ranks_long |>
filter(time == 0) |>
mutate(winf0 = winf) |>
select (trt, id, winf0)

ranks_long <- ranks_long |>
merge(winf0, by = c("trt", "id")) |>
filter(time != "0") |>
arrange(time, id)

######--------------Apply the MMRM to the win fractions -----------------####
# Reorder the levels of 'trt' so that '1' comes before '0'
ranks_long$trt <- factor(ranks_long$trt, levels = c("1", "0"))

mmrm_winf <- mmrm(winf ~ time * trt +winf0*time + winf0 * trt + us(time | trt/id),
data = ranks_long,
control = mmrm_control(method = "Kenward-Roger",
vcov = "Kenward-Roger-Linear"))
summary(mmrm_winf)

# Use emmeans to get the estimated marginal means
emm_results_winf <- emmeans(mmrm_winf, ~ trt | time)

# Perform contrasts for time-specific comparisons of treatment vs control  
contrast_results_winf <- contrast(emm_results_winf, interaction = c("pairwise"))

contrast_summary_winf <- summary(contrast_results_winf , infer = c(TRUE, TRUE))

# Display the estimate, 95% confidence interval, and p-value
print(contrast_summary_winf)

# Calculate WinP and its confidence intervals
mmrm_estimates <- contrast_summary_winf |>
mutate(WinP = estimate / 2 + 0.5,
lgt = log(WinP / (1 - WinP)),
selgt = SE / (WinP * (1 - WinP)),
l = lgt - 1.96 * selgt,
u = lgt + 1.96 * selgt,
low = 1 / (1 + exp(-l)),
upp = 1 / (1 + exp(-u)),
wid = upp - low,
test = lgt / selgt,
p_val = 2 * (1 - pnorm(abs(test))))

# Display WinP estimates, confidence intervals, and p-values
mmrm_estimates |>
select(WinP, low, upp, wid, p_val) |>
print()

 \end{verbatim}
  
  \newpage
   
\nolinenumbers

\bibliography{likertJosh}

 \clearpage
 
\thispagestyle{empty} 
 
\clearpage

\thispagestyle{empty}

\begin{landscape}

\begin{table}

\caption{Percentages of participants who had data deleted (dropout \%) according to trigger value  and probability by treatment arm (control, treatment) and missing data mechanism (missing at random (MAR) and missing not at random (MNAR).}\label{simtable} 

\begin{center}

\begin{tabular}{c|c|c|c|c|c|c|c|c|c|c|c} 
\hline 
 &    &                             & \multicolumn{4}{c}{MAR} & & \multicolumn{4}{c}{MNAR} \\ \cline{4-7} \cline{9-12}

   &    &             & 16, 40\%   &  16,  40\%  &  16, 50\%  &  16, 50\%  & &  16, 40\% & 16, 40\%  & 16,  50\% & 16, 50\%  \\ 

Trajectory & Case  & Group          &  16, 40\% & 15, 40\%   & 16, 30\% & 15, 30\% & & 15, 40\% & 16, 40\%  & 16, 30\% & 15, 30\% \\ 

\hline

1 & 1-4 & 0     &  26 &  19  &  33  & 24 &     &  27 &  20  & 36   & 25\\    \hline
   &  & 1       &  26 & 26   & 20   & 20 &     &   28  & 28 & 23  & 23\\   \hline
 
2& 5-8 & 0     &  30 & 26 & 38 & 33 & &   34 & 29 & 41 & 36\\  \hline
  &   & 1        &  24 & 24 & 19 & 19  &   & 28 & 28 & 22 &22\\   \hline

3 &9-12 & 0      & 26 & 23 & 33 & 29  &   &   29 &25 & 36 & 30\\  \hline
  &   & 1       &    20 &24 & 16 & 19   &   &   22& 27& 18 & 21\\   \hline
 
4& 13-16 & 0     & 22 & 20 & 25 & 25       &     &  25  & 30  &25 & 28 \\    \hline
    &     & 1      &   26 & 26 & 21 & 24      &     &   27 & 27 &  21 & 26 \\   
\hline 

\end{tabular}

\end{center}

\end{table}
  
\end{landscape}

\clearpage

\thispagestyle{empty} 

\begin{table}


 \caption{Performance based on 1000 simulation replicates of  three methods for handling missing data in comparison with no missing data  in estimating landmark  win probability   under data missing completely at random (MCAR) of 30\% in each arm with sample size 50 per group.}\label{mcar}

\begin{center}

\begin{tabular}{c|c|c|c  }
\hline
Trajectory &  Method   & Bias\% $^*$ & \cit{  \rm ML }{\rm CV}{ \rm MR}{\rm WD}{\rm EP}$^\S$ \\
\hline
        1 &  No missing    &    0.2 & \cit{  2.1} { 95.7}{ 2.2} { 22.0}{  4.3} \\ \hline
          &   GPC     &    0.5 & \cit{  2.6} { 94.6}{ 2.8} { 20.5}{  5.4} \\ \hline
         &   CCA     &   -4.5 & \cit{  3.1} { 95.3}{ 1.6} { 26.2}{  4.7} \\  \hline
          &   MMRM    &   -3.2 & \cit{  2.5} { 96.0}{ 1.5} { 25.1}{  4.0} \\  \hline
 
      2 &  No missing    &    0.7 & \cit{  2.2} { 95.6}{ 2.2} { 22.1}{  4.4} \\ \hline
          &   GPC     &   57.5 & \cit{  0.6} { 90.3}{ 9.1} { 20.4}{  9.7} \\ \hline
       &   CCA     &   -4.5 & \cit{  3.1} { 95.3}{ 1.6} { 26.2}{  4.7} \\ \hline
        &   MMRM    &   -3.1 & \cit{  2.5} { 96.1}{ 1.4} { 25.2}{  3.9} \\ \hline
  
       3 & No missing      &   -0.1 & \cit{  2.1} { 95.8}{ 2.1} { 21.8}{ 18.1} \\ \hline
         &   GPC     &  -41.2 & \cit{  7.1} { 92.2}{ 0.7} { 20.4}{ 11.3} \\ \hline
         &   CCA     &   -4.0 & \cit{  2.7} { 95.7}{ 1.6} { 26.0}{ 12.5} \\ \hline
        &   MMRM    &   -3.0 & \cit{  2.4} { 95.9}{ 1.7} { 24.9}{ 14.0} \\ \hline
 
 4 &  No missing      &    0.1 & \cit{  2.0} { 95.9}{ 2.1} { 21.2}{ 55.1} \\ \hline
  &   GPC     & -170.7 & \cit{ 43.1} { 56.8}{ 0.1} { 20.4}{  8.3} \\ \hline
  &   CCA     &   -4.0 & \cit{  3.1} { 95.6}{ 1.3} { 25.4}{ 38.2}  \\ \hline
  &   MMRM    &   -3.4 & \cit{  2.5} { 96.0}{ 1.5} { 24.5}{ 42.8} \\ \hline

\hline
\end{tabular}

\end{center}
 
$^*$ Bias \% is defined as  the mean of  $(\est{\rm WinP} - {\rm WinP})/{\rm WinP}\times 100$ over 1000 simulations.  
$^\S$  Each  95\% confidence interval entry is presented as \cit{ \rm  ML}{\rm CV \%}{\footnotesize\rm MR }{\rm WD \times 100}{ \rm EP}, where ML and MR indicate the percentage of times the upper limit is smaller  and the lower limit is greater than the true parameter value, respectively, CV indicates the percentage of times the confidence interval covers the parameter value, WD is the mean width of 1000 confidence intervals, and EP (empirical power) is defined as the percentage of times that the null hypothesis $H_0:$ WinP=0.5 being rejected at 2-sided 5\% significance level.  


\end{table}

\clearpage
\thispagestyle{empty}

\begin{table}


 \caption{Performance based on 1000 simulation replicates of three methods for handling missing data in comparison with no missing data  in estimating landmark win probability  under data missing at random (MAR) and missing not at random (MNAR) with sample size 50 per group.}\label{traj12}

\begin{center}
\begin{tabular}{c|c|c|c|c|c|c  }
\hline
 Case$^\dagger$              & Method             &      \multicolumn{2}{c}{MAR}  &&  \multicolumn{2}{c}{MNAR} \\ \cline{3-4} \cline{6-7}
   &      & Bias\% $^*$ & \cit{  \rm ML }{\rm CV}{ \rm MR}{\rm WD}{\rm EP}$^\S$ & &Bias & \cit{  \rm ML }{\rm CV}{ \rm MR}{\rm WD}{\rm EP} \\
\hline
 No missing &  Traj 1    &    0.2 & \cit{  2.1} { 95.7}{ 2.2} { 22.0}{  4.3}    & &    0.2 & \cit{  2.1} { 95.7}{ 2.2} { 22.0}{  4.3} \\ \hline
\\
 1  &   GPC     &   36.7 & \cit{  0.8} { 94.2}{ 5.0} { 21.6}{  5.8}    & &   45.1 & \cit{  0.5} { 93.5}{ 6.0} { 20.8}{  6.5} \\  \hline
  &   CCA     &   40.6 & \cit{  1.0} { 92.5}{ 6.5} { 25.8}{  7.5}    & &   46.2 & \cit{  0.7} { 92.8}{ 6.5} { 26.4}{  7.2} \\  \hline
   &   MMRM    &   26.5 & \cit{  1.4} { 94.7}{ 3.9} { 25.5}{  5.3}    & &   38.0 & \cit{  0.7} { 93.8}{ 5.5} { 26.4}{  6.2} \\  \hline
 \\
  
 2 &   GPC     &   30.9 & \cit{  0.8} { 94.6}{ 4.6} { 21.4}{  5.4}    & &   46.1 & \cit{  0.4} { 94.1}{ 5.5} { 20.8}{  5.9} \\  \hline
  &   CCA     &   52.5 & \cit{  0.8} { 91.8}{ 7.4} { 26.1}{  8.2}    & &   70.4 & \cit{  0.2} { 90.4}{ 9.4} { 26.8}{  9.6} \\  \hline
   &   MMRM    &   33.2 & \cit{  1.0} { 94.4}{ 4.6} { 25.6}{  5.6}    & &   56.3 & \cit{  0.3} { 92.3}{ 7.4} { 26.7}{  7.7} \\  \hline
 \\
  
 3 &   GPC     &   51.4 & \cit{  0.5} { 92.1}{ 7.4} { 21.6}{  7.9}    & &   34.3 & \cit{  1.0} { 94.2}{ 4.8} { 20.9}{  5.8} \\  \hline
 &   CCA     &    0.6 & \cit{  2.0} { 95.4}{ 2.6} { 25.8}{  4.6}    & &    0.4 & \cit{  2.3} { 94.7}{ 3.0} { 26.5}{  5.3} \\  \hline
  &   MMRM    &    4.3 & \cit{  2.2} { 95.0}{ 2.8} { 25.6}{  5.0}    & &   -2.6 & \cit{  2.3} { 95.0}{ 2.7} { 26.4}{  5.0} \\  \hline
 \\
 
 4 &   GPC     &   46.6 & \cit{  0.6} { 93.1}{ 6.3} { 21.5}{  6.9}    & &   34.8 & \cit{  0.9} { 94.3}{ 4.8} { 20.9}{  5.7} \\  \hline
   &   CCA     &    6.6 & \cit{  2.0} { 95.3}{ 2.7} { 26.1}{  4.7}    & &   18.1 & \cit{  1.7} { 94.2}{ 4.1} { 26.9}{  5.8} \\  \hline
  &   MMRM    &    7.4 & \cit{  2.0} { 95.1}{ 2.9} { 25.6}{  4.9}    & &   10.5 & \cit{  1.8} { 94.9}{ 3.3} { 26.7}{  5.1} \\  \hline
 \\
 No missing  &   Traj 2   &    0.7 & \cit{  2.2} { 95.6}{ 2.2} { 22.1}{  4.4}    & &    0.7 & \cit{  2.2} { 95.6}{ 2.2} { 22.1}{  4.4} \\  \hline
 \\
 5 &   GPC     &   80.8 & \cit{  0.1} { 87.2}{12.7} { 21.4}{ 12.8}    & &   60.3 & \cit{  0.3} { 92.6}{ 7.1} { 20.9}{  7.4} \\  \hline
   &   CCA     &   16.9 & \cit{  1.5} { 94.1}{ 4.4} { 26.0}{  5.9}    & &   30.2 & \cit{  1.3} { 93.8}{ 4.9} { 27.0}{  6.2} \\  \hline
   &   MMRM    &    4.5 & \cit{  2.0} { 95.0}{ 3.0} { 25.7}{  5.0}    & &   20.3 & \cit{  1.3} { 94.7}{ 4.0} { 27.0}{  5.3} \\  \hline
 \\
  
 6 &   GPC     &   71.3 & \cit{  0.2} { 88.6}{11.2} { 21.4}{ 11.4}    & &   63.9 & \cit{  0.3} { 91.7}{ 8.0} { 20.7}{  8.3} \\   \hline
   &   CCA     &   24.8 & \cit{  1.3} { 94.2}{ 4.5} { 25.6}{  5.8}    & &   47.0 & \cit{  0.6} { 92.5}{ 6.9} { 26.4}{  7.5} \\   \hline
  &   MMRM    &    8.8 & \cit{  1.6} { 95.5}{ 2.9} { 25.3}{  4.5}    & &   34.7 & \cit{  0.9} { 93.7}{ 5.4} { 26.5}{  6.3} \\   \hline
 \\
  
 7 &   GPC     &   96.0 & \cit{  0.1} { 84.5}{15.4} { 21.4}{ 15.5}    & &   47.9 & \cit{  0.5} { 93.4}{ 6.1} { 21.1}{  6.6} \\   \hline
   &   CCA     &  -16.3 & \cit{  3.9} { 94.5}{ 1.6} { 26.2}{  5.5}    & &  -17.9 & \cit{  3.5} { 94.5}{ 2.0} { 27.1}{  5.5} \\   \hline
  &   MMRM    &  -14.0 & \cit{  3.0} { 95.4}{ 1.6} { 25.8}{  4.6}    & &  -22.5 & \cit{  3.5} { 95.0}{ 1.5} { 27.1}{  5.0} \\   \hline
 \\
  
 8 &   GPC     &   85.3 & \cit{  0.2} { 85.6}{14.2} { 21.5}{ 14.4}    & &   51.9 & \cit{  0.5} { 92.9}{ 6.6} { 20.9}{  7.1} \\   \hline
  &   CCA     &   -7.4 & \cit{  2.8} { 95.0}{ 2.2} { 25.7}{  5.0}    & &    1.9 & \cit{  2.2} { 94.5}{ 3.3} { 26.5}{  5.5} \\  \hline
   &   MMRM    &   -8.9 & \cit{  2.9} { 95.4}{ 1.7} { 25.3}{  4.6}    & &   -5.2 & \cit{  2.1} { 95.5}{ 2.4} { 26.5}{  4.5} \\  \hline
\hline
\end{tabular}
\end{center}

$^\dagger$Case number denotes the combination of Trajectory and missing data percentage   as shown in Table \ref{simtable}. 
$^*$ Bias \% is defined as the mean of  $(\est{\rm WinP} - {\rm WinP})/{\rm WinP}\times 100$ over 1000 simulations.  
$^\S$Each  95\% confidence interval entry is presented as \cit{ \rm  ML}{\rm CV \%}{\footnotesize\rm MR }{\rm WD \times 100}{ \rm EP}, where ML and MR indicate the percentage of times the upper limit is smaller  and the lower limit is greater than the true parameter value, respectively, CV indicates the percentage of times the confidence interval covers the parameter value, WD is the mean width of 1000 confidence intervals, and EP (empirical power) is defined as the percentage of times that the null hypothesis $H_0:$ WinP=0.5 being rejected at 2-sided 5\% significance level.  


\end{table}

\clearpage
\thispagestyle{empty} 

\begin{table}


 \caption{Performance based on 1000 simulation replicates of three methods for handling missing data in comparison with no missing data  in estimating win probability at endpoint under data missing at random (MAR) and missing not at random (MNAR) with sample size 50 per group.}\label{traj34}

\begin{center}
\begin{tabular}{c|c|c|c|c|c|c}
\hline
 Case$^\dagger$              & Method             &      \multicolumn{2}{c}{MAR}  &&  \multicolumn{2}{c}{MNAR} \\ \cline{3-4} \cline{6-7}
   &      & Bias\% $^*$ & \cit{  \rm ML }{\rm CV}{ \rm MR}{\rm WD}{\rm EP}$^\S$ & &Bias & \cit{  \rm ML }{\rm CV}{ \rm MR}{\rm WD}{\rm EP} \\
\hline

 No missing &   Traj 3    &   -0.1 & \cit{  2.1} { 95.8}{ 2.1} { 21.8}{ 18.1}    & &   -0.1 & \cit{  2.1} { 95.8}{ 2.1} { 21.8}{ 18.1} \\ \hline
\\
 9 &   GPC     &   21.5 & \cit{  1.6} { 94.7}{ 3.7} { 21.4}{ 25.2}    & &    0.8 & \cit{  2.3} { 94.9}{ 2.8} { 20.6}{ 19.9} \\ \hline
  &   CCA     &   41.7 & \cit{  0.8} { 93.1}{ 6.1} { 25.2}{ 26.3}    & &   42.8 & \cit{  0.5} { 93.5}{ 6.0} { 25.7}{ 23.6} \\ \hline
   &   MMRM    &   28.4 & \cit{  1.1} { 95.2}{ 3.7} { 25.1}{ 21.5}    & &   36.3 & \cit{  0.5} { 94.1}{ 5.4} { 25.7}{ 22.2} \\ \hline
 \\
 
10 &   GPC     &   13.3 & \cit{  2.5} { 94.6}{ 2.9} { 21.3}{ 22.8}    & &   -1.1 & \cit{  2.6} { 95.0}{ 2.4} { 20.6}{ 18.5} \\ \hline
  &   CCA     &   55.0 & \cit{  0.7} { 92.0}{ 7.3} { 25.5}{ 29.7}    & &   65.1 & \cit{  0.3} { 90.9}{ 8.8} { 26.0}{ 30.5} \\ \hline
  &   MMRM    &   35.7 & \cit{  1.0} { 94.7}{ 4.3} { 25.2}{ 23.3}    & &   53.7 & \cit{  0.3} { 92.7}{ 7.0} { 25.9}{ 26.0} \\ \hline
\\
 
11 &   GPC     &   36.8 & \cit{  0.7} { 94.0}{ 5.3} { 21.4}{ 30.5}    & &   -9.4 & \cit{  3.5} { 94.6}{ 1.9} { 20.7}{ 17.4} \\ \hline
  &   CCA     &    1.5 & \cit{  2.5} { 95.1}{ 2.4} { 25.5}{ 15.7}    & &   -2.3 & \cit{  2.7} { 94.9}{ 2.4} { 26.1}{ 13.4} \\ \hline
 &   MMRM    &    5.6 & \cit{  2.1} { 95.8}{ 2.1} { 25.4}{ 16.0}    & &   -3.2 & \cit{  2.5} { 95.2}{ 2.3} { 26.0}{ 12.6} \\ \hline
\\
 
12 &   GPC     &   30.1 & \cit{  1.1} { 94.1}{ 4.8} { 21.3}{ 27.5}    & &  -11.0 & \cit{  3.9} { 94.2}{ 1.9} { 20.7}{ 16.4} \\ \hline
  &   CCA     &    8.7 & \cit{  2.2} { 95.2}{ 2.6} { 25.7}{ 16.8}    & &   14.8 & \cit{  1.5} { 94.7}{ 3.8} { 26.3}{ 15.8} \\ \hline
  &   MMRM    &    9.3 & \cit{  2.2} { 95.4}{ 2.4} { 25.4}{ 16.8}    & &    9.7 & \cit{  1.6} { 94.8}{ 3.6} { 26.2}{ 15.5} \\ \hline
\\

No missing & Traj 4     &    0.1 & \cit{  2.0} { 95.9}{ 2.1} { 21.2}{ 55.1}    & &    0.1 & \cit{  2.0} { 95.9}{ 2.1} { 21.2}{ 55.1} \\  \hline
\\ 
13 &   GPC     &  -89.9 & \cit{ 15.2} { 84.5}{ 0.3} { 21.8}{ 21.0}    & &  -58.2 & \cit{  8.4} { 91.1}{ 0.5} { 20.5}{ 35.3} \\ \hline
 &   CCA     &   69.5 & \cit{  0.4} { 90.3}{ 9.3} { 23.7}{ 69.8}    & &   66.4 & \cit{  0.3} { 91.5}{ 8.2} { 24.2}{ 66.7} \\ \hline
 &   MMRM    &   50.3 & \cit{  0.7} { 92.2}{ 7.1} { 23.6}{ 63.4}    & &   60.5 & \cit{  0.3} { 92.3}{ 7.4} { 24.2}{ 65.4} \\ \hline

\\
 
 14  &   GPC     &  -92.1 & \cit{ 16.2} { 83.6}{ 0.2} { 21.7}{ 20.2}    & &  -71.5 & \cit{  9.7} { 90.0}{ 0.3} { 20.5}{ 31.3} \\ \hline
  &   CCA     &   61.7 &  \cit{  0.7} { 91.2}{ 8.1} { 24.2}{ 65.3}    & &   51.0 & \cit{  0.5} { 92.8}{ 6.7} { 24.8}{ 61.0} \\ \hline
  &   MMRM    &   46.5 & \cit{  0.8} { 92.5}{ 6.7} { 24.0}{ 60.7}    & &   47.4 & \cit{  0.7} { 92.5}{ 6.8} { 24.8}{ 59.0} \\ \hline

\\
 15 &   GPC     &  -71.3 & \cit{ 12.3} { 87.3}{ 0.4} { 21.6}{ 27.2}    & &  -64.7 & \cit{ 10.1} { 89.4}{ 0.5} { 20.5}{ 32.8} \\ \hline
 &   CCA     &   28.7 & \cit{  1.4} { 94.3}{ 4.3} { 24.1}{ 54.4}    & &   20.3 & \cit{  1.7} { 94.6}{ 3.7} { 24.7}{ 48.5} \\ \hline
 &   MMRM    &   25.3 & \cit{  1.4} { 94.8}{ 3.8} { 24.0}{ 53.5}    & &   19.9 & \cit{  1.5} { 95.2}{ 3.3} { 24.7}{ 48.8} \\ \hline

\\
 16 &   GPC     &  -86.7 & \cit{ 13.5} { 86.2}{ 0.3} { 21.6}{ 22.7}    & &  -73.2 & \cit{ 11.1} { 88.4}{ 0.5} { 20.6}{ 30.2} \\ \hline
  &   CCA     &   33.7 & \cit{  1.4} { 93.3}{ 5.3} { 24.3}{ 55.6}    & &   34.8 & \cit{  1.2} { 93.5}{ 5.3} { 24.9}{ 54.5} \\ \hline
  &   MMRM    &   27.2 & \cit{  1.4} { 94.9}{ 3.7} { 24.0}{ 54.2}    & &   30.9 & \cit{  1.2} { 94.0}{ 4.8} { 24.8}{ 53.1} \\ \hline

\hline
\end{tabular}
\end{center}

$^\dagger$Case number denotes the combination of Trajectory and missing data percentage   as shown in Table \ref{simtable}.
$^*$ Bias \% is defined as  the mean of  $(\est{\rm WinP} - {\rm WinP})/{\rm WinP}\times 100$ over 1000 simulations.  
$^\S$Each  95\% confidence interval entry is presented as \cit{ \rm  ML}{\rm CV \%}{\footnotesize\rm MR }{\rm WD \times 100}{ \rm EP}, where ML and MR indicate the percentage of times the upper limit is smaller  and the lower limit is greater than the true parameter value, respectively, CV indicates the percentage of times the confidence interval covers the parameter value, WD is the mean width of 1000 confidence intervals, and EP (empirical power) is defined as the percentage of times that the null hypothesis $H_0:$ WinP=0.5 being rejected at 2-sided 5\% significance level.  


\end{table}

\clearpage
\thispagestyle{empty}

\begin{table}


 \caption{Results of the motivating examples }\label{example}

\begin{center}
\begin{tabular}{c| c | c| c }
\hline
Method &  & WinP estimate (95\% CI) & P-value \\
\hline
\multicolumn{4}{c}{Postnatal depression trial \cite{gregoire96} }  \\   
\\
GPC    & & 0.737 (0.611, 0.834) & 0.0005\\ \hline
\\
CCA   & & 0.779 (0.604, 0.890) & 0.0032 \\ \hline
\\
MMRM & Visit\\
 & 1 &     0.670 (0.516,   0.794) &    0.0314\\
& 2     & 0.700 (0.544,  0.817)  &    0.0132\\
 & 3     & 0.772 (0.619, 0.876)  &     0.0011\\
& 4  &    0.703 (0.521,  0.837) &    0.0300 \\
& 5   &  0.749 (0.583,  0.865) &    0.0048\\
&  6   &  0.774 (0.605, 0.885) &    0.0027 \\
\\
\multicolumn{4}{c}{Labor pain trial \cite{davis1991}}  \\
\\
GPC  & &  0.756 (0.649, 0.838) &   0.000015\\ \hline
\\
CCA & & 0.895 (0.738,  0.962) &   0.00014\\ \hline
\\
MMRM & 30-min interval \\
          & 1 &  0.587 (0.461, 0.702) &   0.1742  \\ \hline
          & 2 &  0.656 (0.527,  0.765) &   0.0182\\ \hline
          & 3 &  0.772 (0.650,  0.861) &   0.0001\\ \hline
          & 4 &  0.844 (0.712,  0.922) &   0.00002 \\ \hline
          & 5 &  0.861 (0.745,  0.930) &    0.00001\\ \hline
          & 6 &  0.875 (0.722,  0.950) &    0.00012 \\ \hline
\hline

\end{tabular}

\end{center}


\end{table}

\end{document}